\documentclass[a4paper,11pt]{article}
\usepackage{pos}

\usepackage{xspace}
\usepackage[utf8]{inputenc}
\usepackage{graphicx}
\usepackage{subcaption}
\usepackage{cancel}
\usepackage{bbm}

\usepackage{color}
\usepackage{dcolumn}   
\usepackage{bm}        
\usepackage{comment}

\DeclareOldFontCommand{\bf}{\normalfont\bfseries}{\mathbf}
\DeclareOldFontCommand{\rm}{\normalfont\bfseries}{\mathrm}

\newcommand{\ee}           {\ensuremath{\mathrm{e^+e^-}}\xspace}

\newcommand{\snn}          {\ensuremath{\sqrt{s_{\mathrm{NN}}}}\xspace}
\newcommand{\pt}           {\ensuremath{p_{\rm T}}\xspace}
\newcommand{\Raa}          {\ensuremath{R_\mathrm{AA}}\xspace}

\newcommand{\Dzero}        {\ensuremath{\mathrm{D^0}}\xspace}

\newcommand{\Dminus}        {\ensuremath{\mathrm{D^-}}\xspace}

\newcommand{\Ds}           {\ensuremath{\mathrm{D_s^+}}\xspace}
 
 \newcommand{\Bplus}           {\ensuremath{\mathrm{B^+}}\xspace}
  \newcommand{\Bc}           {\ensuremath{\mathrm{B^+_c}}\xspace}

\newcommand{\Lc}           {\ensuremath{\Lambda_\mathrm{c}^+}\xspace}

\newcommand{\XicPlusZero}  {\ensuremath{\Xi_\mathrm{c}^{+,0}}\xspace}
\newcommand{\Omegac}       {\ensuremath{\Omega_\mathrm{c}^0}\xspace}

\newcommand{\Lb}        
{\ensuremath{\Lambda_\mathrm{b}^0}\xspace}

\newcommand{\LcD} {\ensuremath{\Lc/\Dzero}\xspace}

\title{Heavy flavors and quarkonia: highlights, open questions, and perspectives}

\author*[a]{Andrea Dubla}

\affiliation[a]{GSI Helmholtzzentrum f{\"u}r Schwerionenforschung, 64291 Darmstadt, Germany}

\emailAdd{a.dubla@gsi.de}

\abstract{An overview of the phenomenology and experimental results on open
heavy-flavour and quarkonium production in heavy-ion collisions at the RHIC and at the LHC energies is presented, with special emphasis on observables that carry information from the different collision stages. Perspective for future measurements and phenomenological modeling, that will shed light on the current open question in heavy-ion collisions, will be also discussed.}

\FullConference{HardProbes2023\\
 26-31 March 2023\\
 Aschaffenburg, Germany\\}

\begin{document}
\maketitle

\section{Introduction}

\noindent Heavy-flavour hadrons, containing a charm $(c)$ or beauty $(b)$ quark and a light quark, and quarkonia, bound states of a $c\overline{c}$ or $b\overline{b}$ pair, are effective probes of the properties of the hot and dense strongly-interacting medium formed in high-energy heavy-ion collisions.
Heavy quarks are produced in the early stage of the collision in primary partonic scatterings with large virtuality $Q$ and, thus, on temporal and spatial scales, $\Delta \tau \sim \Delta r \sim 1/Q$, which are sufficiently small for the production to be unaffected by the properties of the medium, in the case of nucleus–nucleus collisions. 
In contrast to light quarks and gluons that can be produced or annihilated during
the entire partonic phase of the medium, heavy quarks preserve their flavour and mass identity while traversing the medium and can be tagged via the measurement of heavy-flavour hadrons in the final state of the collision. This feature makes heavy-flavour hadrons and their decay products effective probes to study and characterise all phases of a heavy-ion collision. In the next sections a selection of open heavy-flavour and quarkonia results, connected to different collision stages, will be discussed together with future perspectives that will allow drawing more firm conclusions.

\section{Initial and pre-equilibrium phases}

\noindent The initial stage of ultra-relativistic heavy-ion collisions with a non-zero impact parameter are characterized by extremely strong electromagnetic fields ($10^{18}-10^{19}$ Gauss) primarily induced by spectator protons and angular momentum ($L \sim \mathcal{O}(10^7 \hbar)$).
The time-dependent magnetic field and the expanding QCD matter induce charged currents due to the combination of Lorentz and Coulomb forces. In an electrically conducting plasma, the induced charged currents might slow down the decay of the magnetic field.
The charm quark is an optimal candidate to probe the properties of this magnetic field because its formation timescale is comparable to the timescale when the magnetic field attains its maximum value and the kinetic relaxation time of charm is similar to the QGP lifetime. These factors allow the charm quark to retain the initial kick picked up from the electromagnetic field, resulting in a significantly larger signal compared to that of lighter quarks.

\begin{figure}[ht!]
    \begin{center}
    \includegraphics[width = 0.45\textwidth]{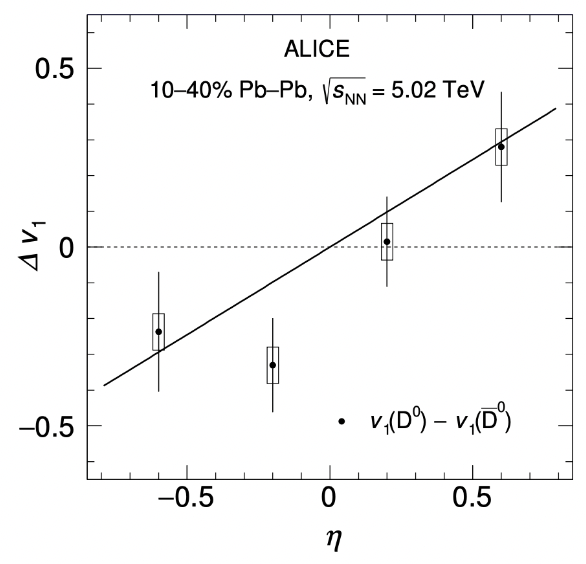}  
    \includegraphics[width = 0.5\textwidth]{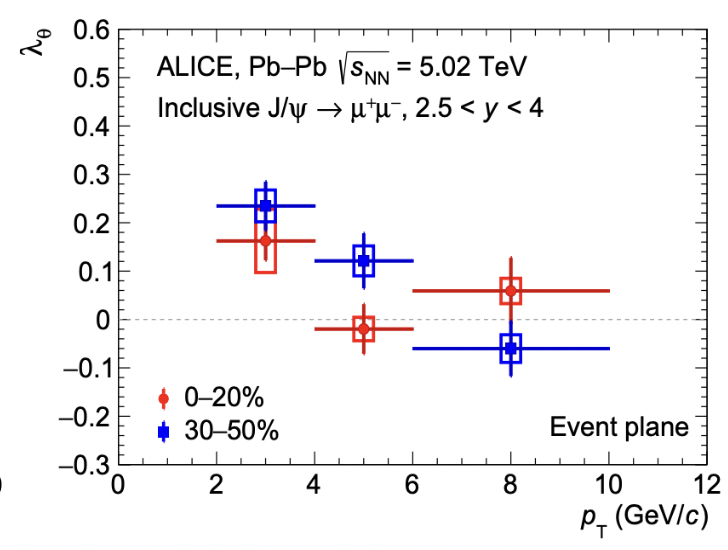}
    \end{center}
    \caption{Left panel: $\Delta v_1$ for \Dzero meson as
a function of pseudorapidity for 3 $< \pt >$ 6 GeV/$c$ for 5--40\% centrality Pb--Pb collisions at
$\snn$ = 5.02 TeV. Right panel: \pt~dependence  of the $\lambda_\theta$ parameter measured with respect to the event plane direction in Pb--Pb collisions at $\snn$ = 5.02 TeV.}
    \label{fig1}
\end{figure}

\noindent One of the observables that can be used to calibrate the strength and lifetime of the magnetic field is the charge dependence of the produced particle directed flow relative to the spectator plane, the so called $\Delta v_1$ \cite{Dubla:2020bdz}. 
Experimental results at the RHIC by the STAR Collaboration and at the LHC by the ALICE Collaboration have shown indications for a different magnitude of the signal and an opposite slope of the $\Delta v_1$ \cite{ALICE:2019sgg,STAR:2019clv}. In the left panel of Fig.~\ref{fig1} the $\Delta v_1$ for $\Dzero$ meson as a function of the pseudorapidity measured by the ALICE Collaboration is shown. The dependence of the charge splitting on collision energy is predicted to be mostly flat and negative~\cite{Chatterjee:2018lsx}, in contrast to the experimental findings. The opposite sign of the measured $\Dzero$ meson $\Delta v_1$ slope at the LHC with respect to model calculations might indicate a stronger effect of the Lorentz force relative to the Coulomb one. However, the experimental measurements are not yet fully significant due to the large uncertainties and did not allow for drawing firm conclusions. Future and more precise measurements, thanks to the much larger statistics that will be collected in Run 3 at the LHC, will shed light on these tantalising effects and differences.

\noindent Another observable that is used to investigate such strong magnetic fields, as well as the huge orbital angular momentum of the medium is the polarisation of quarkonia which can be altered because of these specific features of the environment. The ALICE Collaboration reported the first measurement of the J/$\psi$ polarisation in Pb–Pb collisions at $\snn$ = 5.02 TeV with respect to an axis perpendicular to the event plane \cite{ALICE:2022sli}. By measuring the polarisation with respect to the
estimated reaction plane of the nuclear collision one selects a reference frame that should naturally be connected with the observation of polarisation effects due to the presence of early electromagnetic fields and/or QGP vorticity. These effects can be studies measuring the $\rho_{00}$ element of the spin-density matrix or via the related $\lambda_\theta$ parameter.
The results as a function of $\pt$ for central and semi-central centrality intervals are reported in the right panel of Fig.~\ref{fig1}. A small but significant polarisation effect ($\lambda_\theta > 0$), reaching 3.9$\sigma$ for 2 $< \pt <$ 4 GeV/$c$ and 30–50\% centrality, is measured.
The results correspond to inclusive J/$\psi$ production, implying that a small contribution from a potential polarisation of parent beauty hadrons, which could anyway be diluted in the decay process, might be present. However, a solid interpretation of the result that would allow a quantitative understanding of this observation and a precise connection with the QGP properties at its origin requires dedicated theory studies, which unfortunately are not yet available in the field.  From the experimental side, future polarisation studies should be performed using the $\rm D^{*+}$ vector meson, which could bring additional information about the charm spin alignment expected at the very early stages of heavy-ion collisions due to the produced magnetic field, transferred to the $\rm D^{*+}$ mesons via the hadronisation. 

\section{Quark--gluon plasma phase}

\begin{figure}[ht!]
    \begin{center}
    \includegraphics[width = 0.475\textwidth]{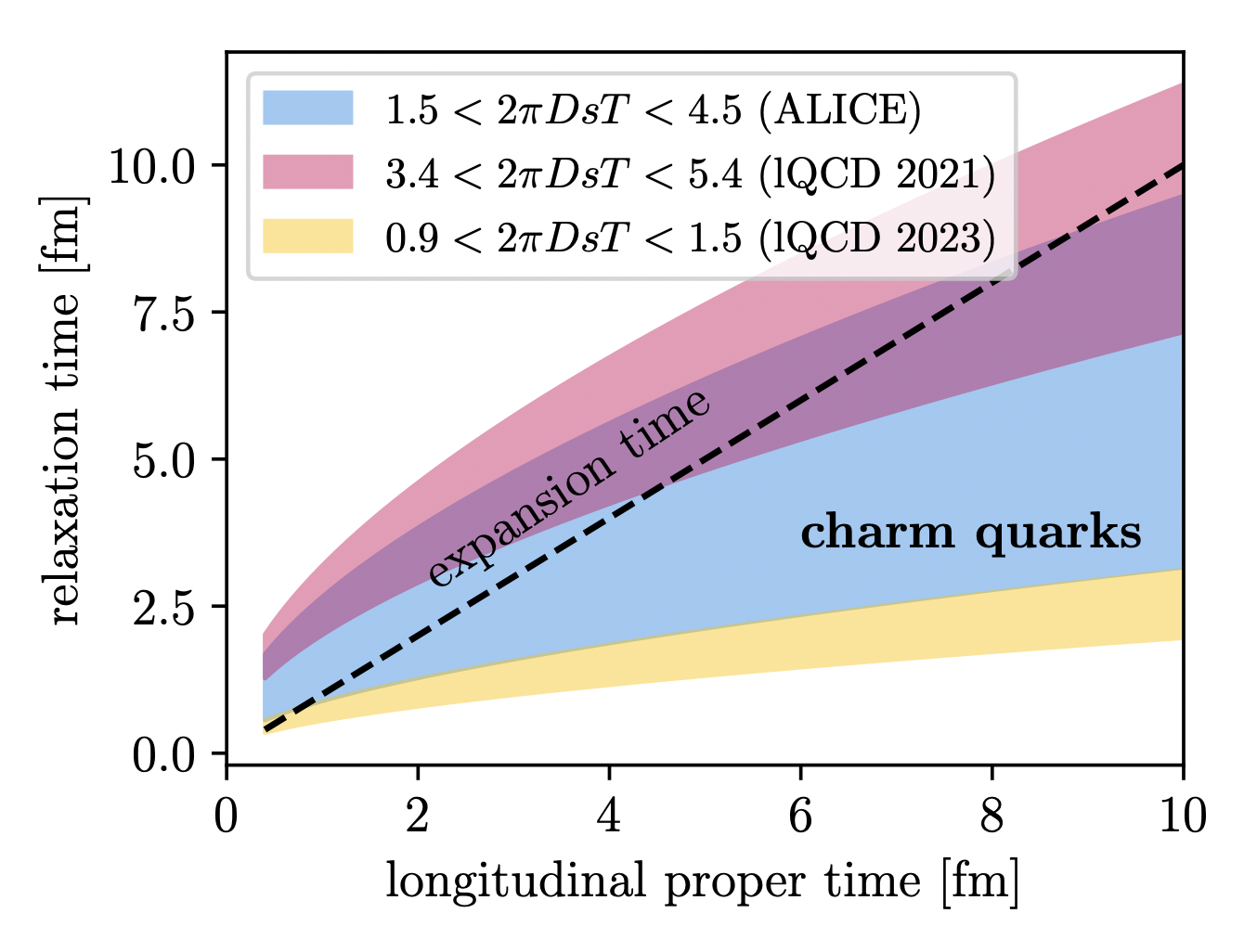}
    \includegraphics[width = 0.48\textwidth]{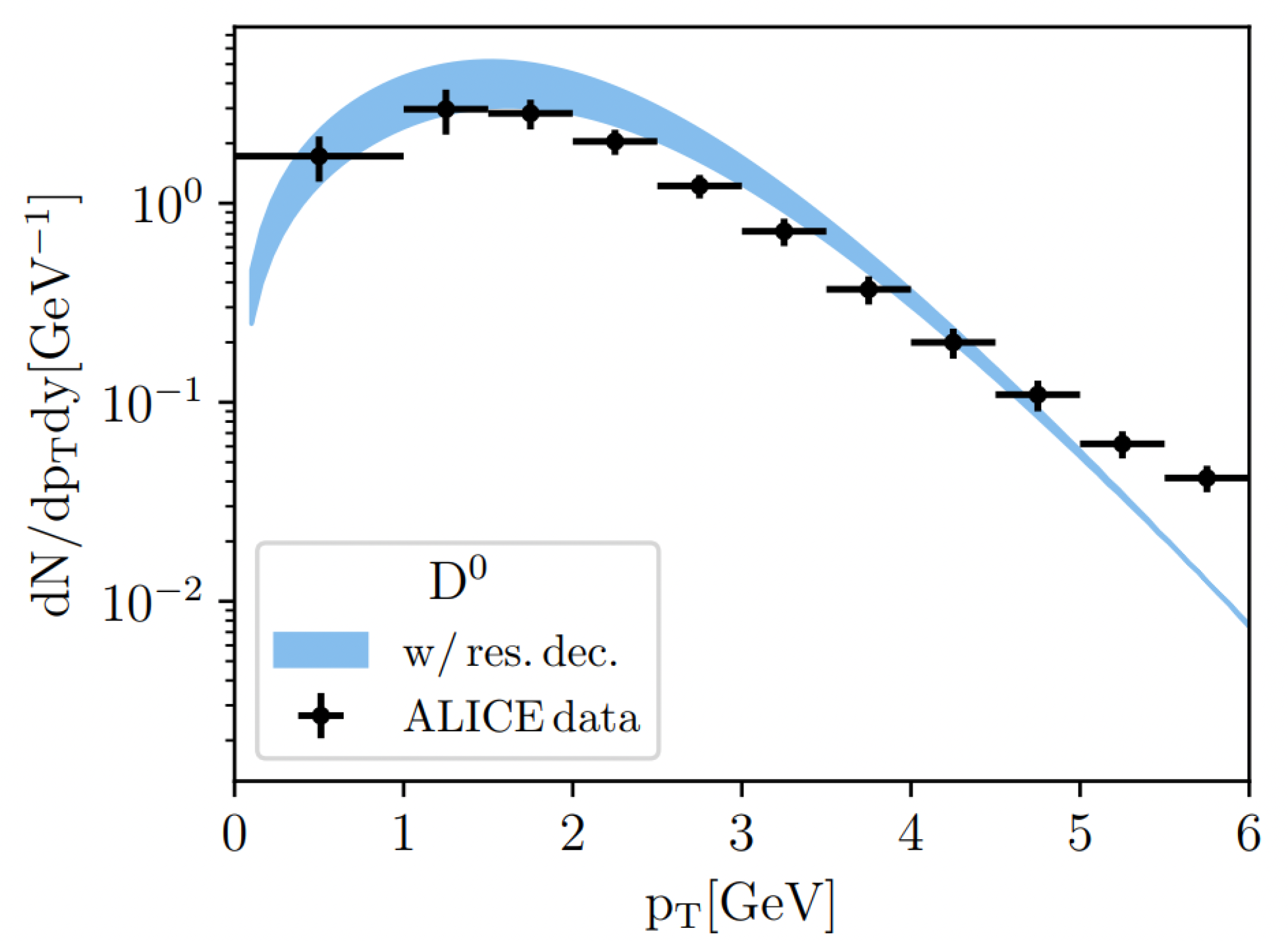}  
    \end{center}
    \caption{Left panel: the relaxation time $\tau_n$ of charm quarks as a function of the longitudinal proper time is compared to the typical expansion timescale $\tau_{\rm exp}$ of the fluid undergoing a Bjorken flow. Right panel: transverse momentum distribution of $\Dzero$ mesons obtained via a fluid-dynamic approach compared with the ALICE result.}
    \label{fig2}
\end{figure}

\noindent Charm and beauty quarks, which are produced via hard scattering processes and have a much larger mass than light quarks, are suitable probes to describe the QGP. Since they are expected to take longer to approach local kinetic equilibrium with respect to the light degrees of freedom, their in-medium dynamics remain a challenging topic. However, recent experimental measurements have shown how J/$\psi$ and D mesons display a strongly positive elliptic flow, suggesting an early thermalization of charm quarks within the QGP. The idea of charm thermalization was already suggested in the context of the Statistical Hadronization Model for charm SHMc \cite{Andronic:2021erx} and recent lattice-QCD (lQCD) calculations support it \cite{Altenkort:2023oms}. In a recent work \cite{Capellino:2022nvf}, the question of charm thermalization was tackled from a dynamical point of view, studying the hydrodynamization time of charm quarks in the context of an expanding medium. 
In Fig.~\ref{fig2} the comparison between the expansion time ($\tau_{\mathrm{exp}}$) and the relaxation time ($\tau_n$) as functions of the longitudinal proper time is reported for the charm quark. This is done assuming an initial temperature of 0.45 GeV, initialization time $\tau_0 = 0.5$ fm/c and employing different values of the diffusion coefficient $D_s$. It is shown that $\tau_n$ goes below $\tau_{\mathrm{exp}}$ quite fast when using transport coefficients arising from fits to experimental data, indicating that the conditions for a fluid-dynamic description are fulfilled for a sizeable fraction of the deconfined fireball lifetime. The most recent lQCD calculation of $D_s$ suggests even an early hydrodynamization time-scale \cite{Altenkort:2023oms}, which is shorter than the typical expanding time scale of the medium.

\noindent This result motivated the derivation of a fluid-dynamic description of charm quark \cite{Capellino:2023cxe}. In the right panel of Fig. \ref{fig2}, the transverse momentum distribution for $\Dzero$ meson is shown in comparison with the ALICE measurement. This fluid-dynamic approach seems to capture the physics description up to $\pt \sim$ 4--5 GeV/$c$, confirming that even the charm quark gets thermalized. The uncertainty band of the model reflects the usage of different $2\pi T D_s$ values ranging from 0 to 1.5. 

\begin{figure}[ht!]
    \begin{center}
    \includegraphics[width = 0.52\textwidth]{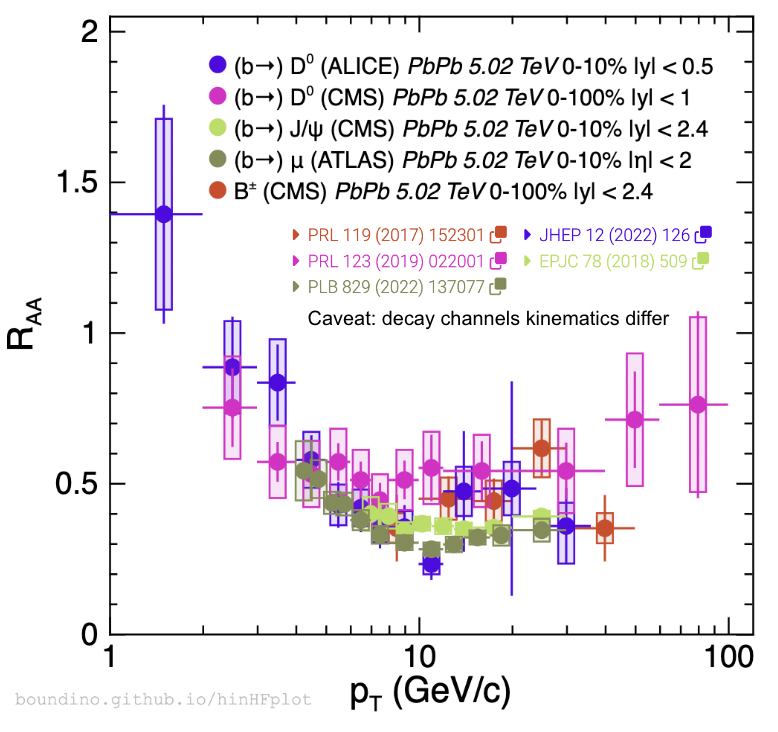}
    \includegraphics[width = 0.46\textwidth]{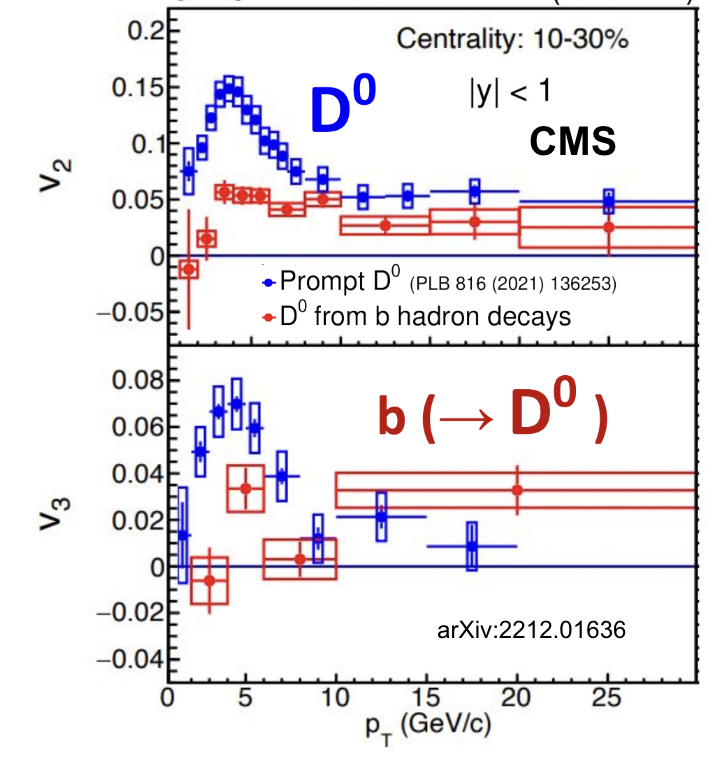}  
    \end{center}
    \caption{The $\Raa$ (left panel) and the $v_2$ and the $v_3$ flow coefficients (right panels) of $\Bplus$ meson and hadrons coming from beauty hadron decays.}
    \label{fig3}
\end{figure}

\noindent Regarding beauty quarks, $D_s$ estimates predict the hydrodynamization time-scale might be of the order of the typical lifetime of the QGP and might not fulfill the conditions for fluid-dynamic applicability. However, the most recent lQCD calculations suggest a possible partial hydrodynamization of beauty quarks.
From the experimental side, several measurements of open and hidden beauty have been published, however, the precision achieved does not yet allow us to draw firm conclusions on their dynamic in the QGP. In the right panel of Fig. \ref{fig3} a compilation of \Raa measurements related to beauty quark from ALICE, CMS, and ATLAS are reported \cite{CMS:2017uoy,ATLAS:2021xtw,ALICE:2022tji}. They demonstrated that energy loss is the dominant mechanism responsible for the measured suppression for $\pt >$ 4 GeV/$c$. It was also shown that the non-prompt
$\Dzero$ $\Raa$ is systematically higher than the prompt $\Dzero$ one for $\pt >$ 5 GeV/$c$ in both central and semi-central centrality classes, indicating that non-prompt $\Dzero$ mesons are less suppressed than prompt $\Dzero$ ones and supporting the expectation that beauty quarks lose less energy than charm quarks because of their larger mass. 
The CMS Collaboration also released anisotropic flow measurements showing non-zero values of elliptic flow of non-prompt \Dzero mesons \cite{CMS:2022vfn}. The $v_2$ coefficient (shown in the right panel of Fig. \ref{fig3}) is significantly lower than in the prompt \Dzero meson case, indicating a different degree of participation to the collective motion of the medium between charm and beauty quarks. The non-prompt $v_3$ coefficient results have large statistical uncertainties so that neither the $\pt$ nor the centrality dependence can be determined.
However, results of $\Raa$ and $v_2$ are still not conclusive when the physics of low $\pt$ beauty quark tries to be addressed, making future Run 3 measurements extremely important to further address the dynamic of the beauty quark and to further constrain QGP parameters using heavy quarks as probes.

\section{Hadronisation mechanisms}

\begin{figure}[ht!]
    \begin{center}
    \includegraphics[width = 0.47\textwidth]{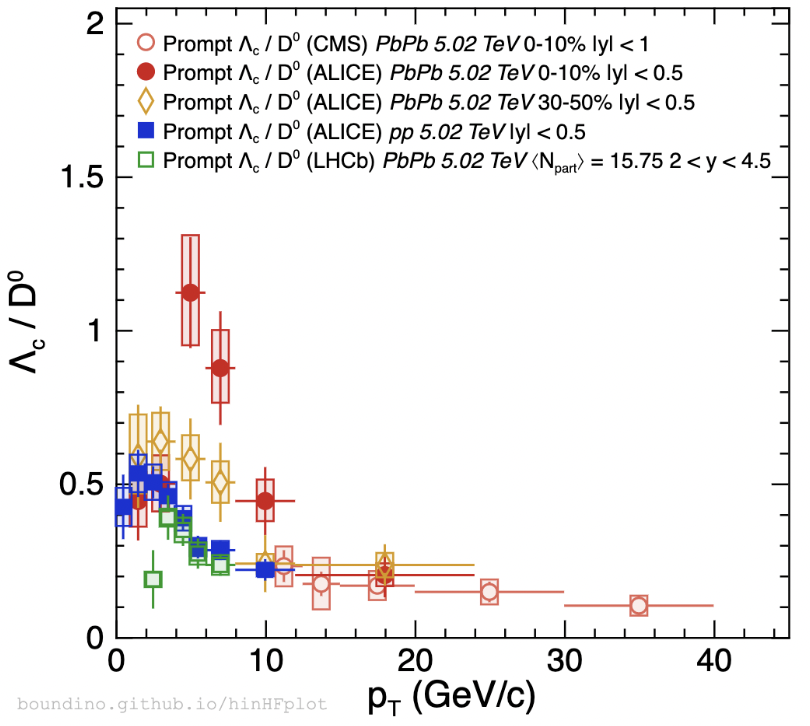}
    \includegraphics[width = 0.5\textwidth]{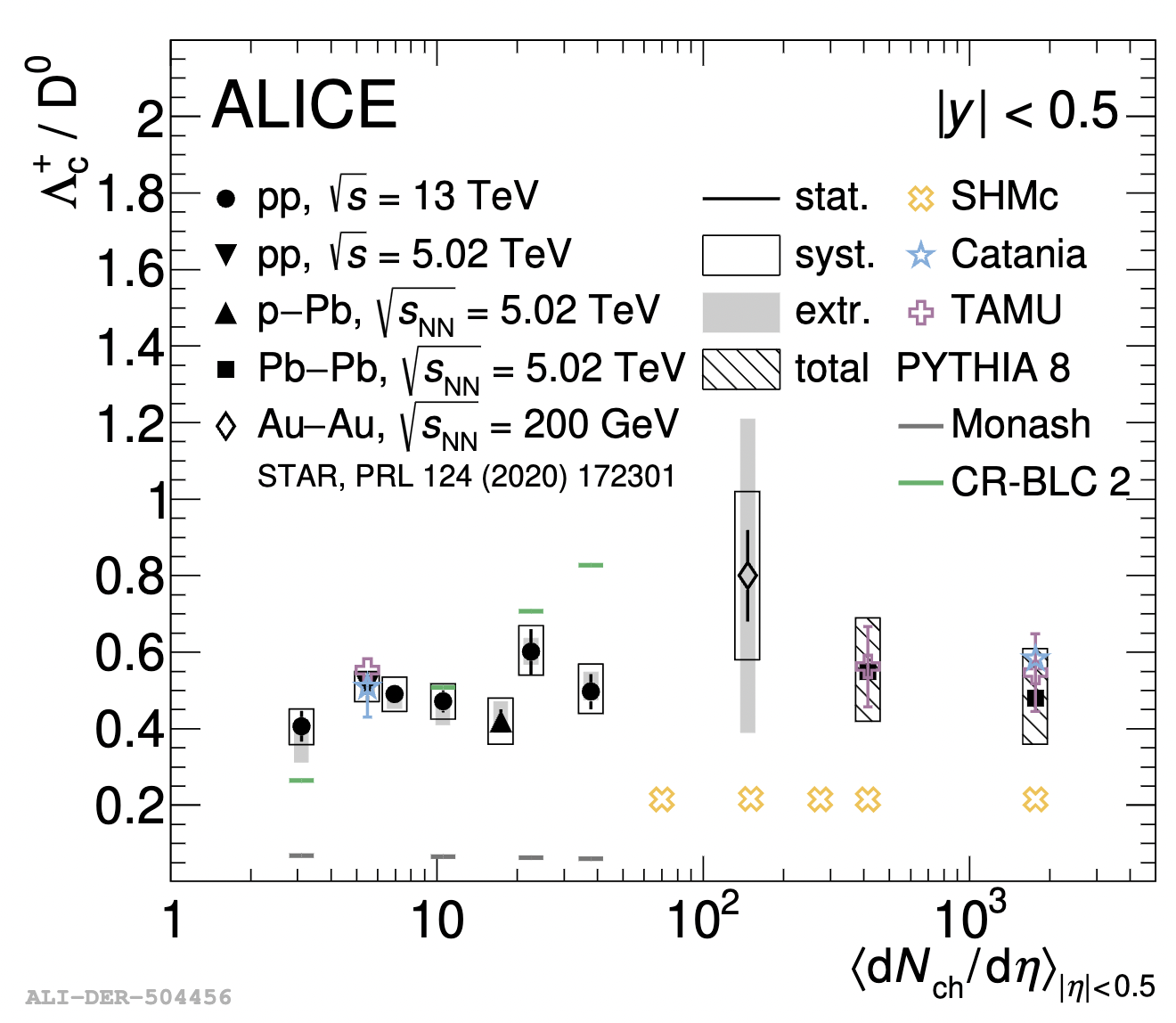}  
    \end{center}
    \caption{Left: comparison of the $\LcD$ cross-section ratios measured in pp collisions by ALICE and in Pb–Pb collisions by ALICE, CMS, and LHCb in different centrality intervals. Right: $\LcD$ cross-section ratio $\pt$-integrated as a function of charged-particle multiplicity at midrapidity measured by ALICE in pp, p–Pb, and Pb–Pb collisions and by STAR in Au–Au collisions.}
    \label{fig4}
\end{figure}

\noindent Measurements of hadron-to-hadron production cross-section ratios in hadronic collisions at LHC energies are sensitive to fragmentation fractions and heavy flavour hadronisation mechanisms.
Measurements of open charm- and beauty-meson production in pp collisions are successfully described by quantum chromodynamics (QCD) calculations based on the factorization of the soft (non-perturbative) and hard (perturbative) processes, where the fragmentation process can be parametrized from measurements performed in \ee or ep collisions, assuming that the hadronisation of charm and beauty quarks is independent of the collision system. 
In the left panel of Fig.~\ref{fig4}, the evolution of the $\pt$-differential $\Lc/\Dzero$ is shown, by comparing the measurements performed in minimum-bias pp collisions, and in different centralities in Pb--Pb collisions at the LHC~\cite{ALICE:2021bib,LHCb:2022ddg,Sirunyan:2019fnc}. At intermediate $\pt$, the ratio increases with centrality, with LHCb data at forward rapidity overlapping with ALICE pp data at midrapidity.
Several models based on different developments and assumptions, like the inclusion of string formation beyond the leading-colour approximation~\cite{Christiansen:2015yqa}, or the inclusion of hadronisation via coalescence~\cite{Minissale:2020bif}, or considering a set of yet-unobserved higher-mass charm-baryon
states~\cite{He:2019tik}, have been proposed to explain the observed experimental baryon enhancement in the different collision systems. 

\noindent The right panel of Fig.~\ref{fig4} shows the evolution of the \pt-integrated $\Lc/\Dzero$ ratio as a function of the charged-particle multiplicity at midrapidity measured by ALICE in pp, p--Pb, and Pb--Pb collisions~\cite{ALICE:2021bib} and by STAR in Au--Au collisions~\cite{STAR:2019ank}. The ratio is approximately around 0.5, thus much higher than the $\approx 0.11$ value measured in $\ee$ and ep collisions, and, within uncertainties, it does not show a dependence on multiplicity. The same ratio in pp (minimum bias) and in Pb--Pb collisions is expected by the Catania and TAMU models, which reproduce also quantitatively the data. The SHMc model also does not expect a dependence on centrality in Pb--Pb collisions. However, it predicts a ratio of around 0.2 and underestimates the data. It must be noted that SHMc does not include the augmented set of high-mass baryon states considered by TAMU. In pp collisions, the PYTHIA 8 Monash tune strongly underestimates the data and expects a slightly decreasing trend with multiplicity. On the contrary, PYTHIA 8 with CR-BLC expects a significant increase of the ratio with multiplicity, deriving from the larger probability of forming junctions colour connections in events with a higher number of MPI, which is not supported by the data. The region of very low multiplicities is not covered by measurements, thus leaving open the question if the $\ee$ values might be recovered when the hadronic activity in the event is very small. 
These heavy-flavour baryon measurements challenge the assumption of that heavy quark hadronisation is a universal process across different colliding systems. The $\Lc$ is the only heavy-flavour baryon whose production was measured in nucleus--nucleus collisions and it will be of primary importance in the near future to perform experimental measurements of additional charm and beauty baryons ($\Lb$, $\XicPlusZero$, and $\Omegac$) in order to further constrain model calculations based on different assumptions.
Additional possible windows to study hadronization of heavy-flavour in hadronic collisions consist in performing precise measurements of  the $\Bc$ meson, which contains a $b$ and a $c$ quarks, multi-charm baryons, such as the $\Xi^+_{\rm cc}$, $\Xi^{++}_{\rm cc}$, $\Omega^+_{\rm cc}$, $\Omega^{++}_{\rm ccc}$, and exotica like the X(3872) and the $\rm T^+_{cc}$.

\section{Hadronic phase and rescattering}

\begin{figure}[ht!]
    \begin{center}
    \includegraphics[width = 0.48\textwidth]{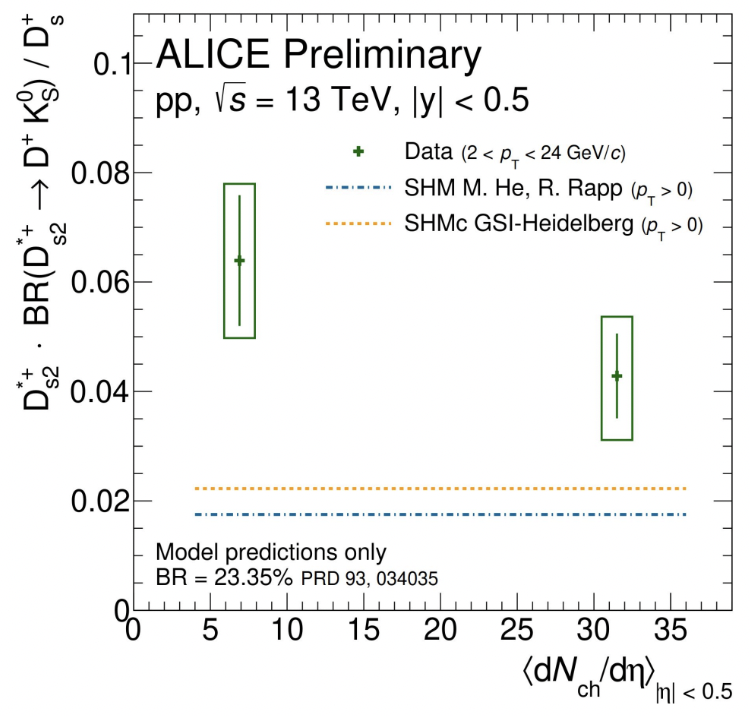}
    \includegraphics[width = 0.5\textwidth]{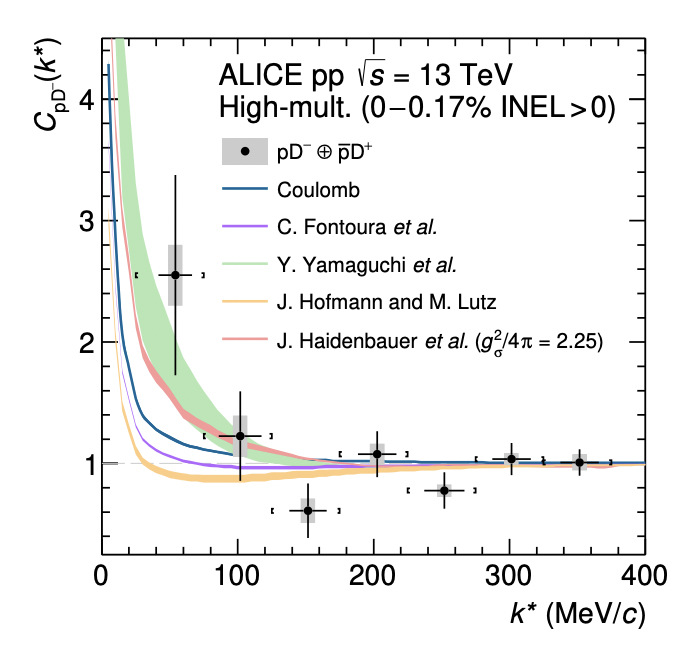}  
    \end{center}
    \caption{Left panel: $\pt$-integrated yields times the BR of the analyzed decay channel of prompt $\rm D^{*+}_{s2}$ mesons divided by the $\pt$-integrated yields of prompt $\Ds$ mesons as a function of the charged-particle multiplicity. Right panel: genuine p-$\Dminus$ correlation function compared with different theoretical models.}
    \label{fig5}
\end{figure}

\noindent The hadron-gas phase lasts approximately 5-10 fm/$c$ from the chemical freeze-out, right after the hadronisation of the QGP, to the kinetic freeze-out, when all interactions cease and hadrons stream freely. The resonances with lifetimes of the same timescale are likely to decay before the kinetic freeze-out. They are therefore good probes of the dynamics of the hadronic phase. In fact, the decay products of resonances are subject to elastic interactions in the hadron gas, which modify their momenta and prevent the reconstruction of the resonance signal by means of an invariant mass analysis. As a consequence, the measured resonance yield is suppressed with respect to the amount produced at the chemical freeze-out. ALICE recently performed the first measurement of $\rm D^+_{s1}(2536)$ and $\rm D^{*+}_{s2}(2573)$ resonances in pp collisions. While no clear multiplicity dependence is observed for the $\rm D^+_{s1}(2536)/\Ds$ ratio, the $\rm D^{*+}_{s2}(2573)/\Ds$ result shows a hint of decreasing trend (left panel of Fig.~\ref{fig5}), though with large uncertainties. If confirmed by more precise measurements, such results might shed light on the interplay between the different hadron lifetimes and the hadronic rescattering, which is expected to influence their abundance.

\noindent The study of the residual strong interaction among hadrons is a very active field within nuclear physics. This interaction can lead to the formation of bound states, such as nuclei, or molecular states.
The study of interactions involving charm mesons has gained significant interest after the observation of the charm-strange meson $\rm D^*_{s0}(2317)$, whose mass lies significantly below the quark model predictions, preventing its accommodation in simple constituent quark models. Also other exotic states, like the X(3872), the $\rm T_{cc}^+$, cannot yet be unambiguously interpreted. The current models, lacking any experimental constraints about the charm hadrons strong
interaction, are not yet able to provide a solid answer about the nature of exotic objects leaving different
interpretations, like molecular states, orbital excited resonance states, and tetraquarks/pentaquarks,
still open.
The knowledge of the charm-hadron interactions is essential also for the study of ultrarelativistic heavy–ion collisions. These interactions modify the heavy-ion observables, and, to disentangle this effect from the signatures of the QGP formation, the scattering parameters of the charm hadrons with light-flavor hadrons, mainly pions and kaons, are required. 
Most recently the first measurement of D-p correlation function was performed within the ALICE Collaboration \cite{ALICE:2022enj}. The result is reported in the right panel of Fig.~\ref{fig5}  and it is compared to several model calculations. The D-p correlation function does not yet provide the precision necessary to distinguish among the different models, shown by the colored bands. However, the result suggests a shallow attractive interaction
with the possible formation of ND bound states. The ALICE Collaboration most recently also released preliminary results for the DK and D$\pi$ pairs. The DK results are in agreement with theoretical predictions based on quantum chromodynamics calculations on the lattice and on chiral effective field theory. In the case of the D$\pi$ interaction instead the measured scattering parameters are significantly smaller than those predicted by the theory. The extracted scattering parameters for both $\rm D^*\pi$ and the $\rm D^*K$ interactions are compatible with zero within the large uncertainties.

\begin{figure}[ht!]
    \begin{center}
    \includegraphics[width = 1.\textwidth]{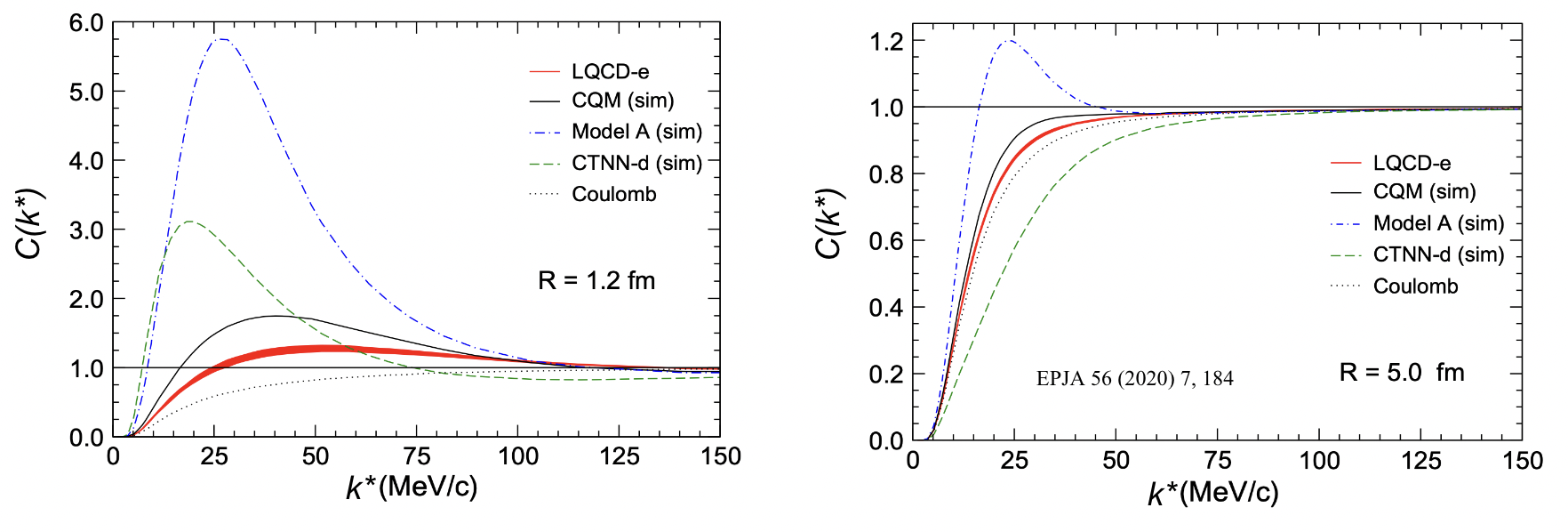}
    \end{center}
    \caption{Predictions for the $\Lc$-p correlation functions including the Coulomb interaction for two different source radii for several model calculations}
    \label{fig6}
\end{figure}

\noindent The little experimental knowledge available about the charm hadron's strong interactions hinders progress in the long-standing issue regarding the existence of charm hypernuclei. A first measurement of $\Lc$-p interactions would be important to calibrate the strength of the potential responsible for binding the $\Lc$ baryon to ordinary matter.
The prospects for deducing constraints on the interaction of charmed baryons with nucleons from experimental measurements were recently studied by computing the $\Lc$-p correlation functions obtained from lQCD and phenomenological potentials available in the literature \cite{Haidenbauer:2020kwo}. In general, the phenomenological models suggest a more strongly attractive $\Lc$-p force than lQCD, and some even lead to two-body bound states. The bound states predicted survive despite of the Coulomb repulsion. The correlation functions computed using the \Lc-p potential from lattice results and by the other phenomenological models are reported in Fig.~\ref{fig6}. In the left panel, the correlation function is produced for a source size of R = 1.2 fm, which is compatible with the one produced in pp collisions. In the right panel, the correlation function is computed for a source size R = 5 fm, which corresponds to the source size produced in heavy-ion collisions. Already at first sight it is clear that the different potentials considered lead to quite different predictions for the $\Lc$-p correlation functions. Specifically, more attractive interactions yield to larger values of the correlation function. 
The presence of a repulsive Coulomb force in the \Lc-p system leads to a strong depletion of the correlation function for small momenta. As a consequence, the signal due to the strong interaction is significantly reduced. Nonetheless, at least for pp collisions with source radii around 1.2 fm the effect by the $\Lc$-p interaction will be detectable in an experiment. For heavy-ion collisions with a typical source radius of around 3-5 fm it looks more challenging. An interesting behaviour is shown by the predictions of the CTNN-d model, which supports bound states. The delicate interplay between the repulsive Coulomb interaction and the strongly attractive \Lc-p potential produces a distinct dependence on the source radius and this characteristic behaviour constitutes an extremely useful signature that will help for either confirming or ruling out such bound states in experiments. Even weakly attractive forces such as those suggested by present-day lattice simulations lead to effects that will be detectable in experiments. It is safe to say that even an experiment with moderate statistics will be sufficient to
discriminate among the calculations.

\bibliographystyle{JHEP}
\bibliography{bibliography}

\end{document}